\documentclass[preprint,preprintnumbers,amsmath,amssymb,12pt]{revtex4}

\usepackage{graphicx}
\usepackage{epsfig}
\usepackage{bm}
\usepackage{amsfonts}
\usepackage{multirow}
\usepackage{float}
\usepackage{xcolor,soul}
\setul{-.6ex}{2pt}
\setstcolor{red}

\usepackage{array}
\newcolumntype{C}[1]{>{\centering\arraybackslash}p{#1}}

\usepackage{subcaption}
\usepackage{hyperref}
\hypersetup{colorlinks=true,citecolor=blue}

\graphicspath{{./figs/}}

\sethlcolor{green}

\begin{document}

\title{Fractional power law inflationary potentials in loop quantum cosmology with inverse volume corrections in light of ACT observations}

\author{Farough Parvizi\footnote{farough.parvizi@uok.ac.ir} and Kayoomars Karami\footnote{kkarami@uok.ac.ir}}
\affiliation{\small{Department of Physics, University of Kurdistan, Pasdaran Street, P.O. Box 66177-15175, Sanandaj, Iran}}

\date{\today}

\begin{abstract}
Here, we investigate the observational viability of fractional power law inflationary potentials, $V(\varphi) \propto \varphi^n$ ($n=$ 1/3, 2/5, 2/3), within the effective framework of Loop Quantum Cosmology (LQC) incorporating inverse volume corrections. By employing LQC modifications to the background dynamics and cosmological perturbation equations in the semi-classical regime, we analytically derive the scalar spectral index $n_{\rm s}$ and the tensor-to-scalar ratio $r$. These theoretical predictions are confronted with the latest high precision joint observational constraints, including ACT DR6, Planck 2018, DESI BAO, and BICEP/Keck datasets (P-ACT-LB-BK18). While steeper classical power law models are in severe tension with modern data, our results demonstrate that the inclusion of LQC inverse volume effects induces a prominent negative shift in $n_{\rm s}$. Consequently, for specific viable ranges of the quantum geometric parameters ($\sigma$ and $\delta$), the theoretical predictions are translated horizontally across the $r-n_{\rm s}$ plane. This mechanism successfully steers classically disfavored models back into the tightly constrained 68\% and 95\% CLs, significantly improving their consistency with precision cosmological data.

\end{abstract}


\maketitle

\section{Introduction}
Cosmic inflation, an idea proposed by Alan Guth in the early 1980s, not only resolves several classical problems of cosmology (such as the flatness, horizon, and monopole problems) but also describes a phase of accelerated and extremely rapid expansion of the universe during a fraction of a second after the Big Bang \cite{Guth:1981, Linde:1982}. According to this theory, quantum fluctuations of the inflaton field are stretched to cosmological scales due to exponential expansion, ultimately leading to the formation of large scale structures of the universe \cite{Bauman:2009}.
While the overarching inflationary paradigm has been extraordinarily successful, identifying the specific underlying microphysics that drove this expansion remains a phenomenological challenge. The literature contains a wide variety of proposed inflationary potentials, many of which are now being tightly constrained or entirely ruled out by increasingly precise Cosmic Microwave Background (CMB) measurements \cite{Planck:2018inflation,Martin:2013tda}. Consequently, current observational data plays a crucial role in breaking the degeneracy among these models, forcing us to continuously refine or discard specific potentials to match the empirical bounds.

Over the past few decades, progressive measurements of CMB anisotropies have strictly tested these theoretical frameworks. Most recently, joint observational constraints from the Atacama Cosmology Telescope (ACT) DR6, combined with the Planck satellite, DESI BAO, and BICEP/Keck (P-ACT-LB-BK18), have significantly updated the phenomenological landscape. These measurements reveal that the best fit value of the scalar spectral index, $n_{\rm s}$, has shifted toward larger values. Because of this shift, several previously favored potentials, such as the Starobinsky model, now exhibit considerable tension with the data. While completely consistent with earlier observations, their predictions are now pushed to the boundary of the 95\% CL. This rising tension has motivated intense theoretical effort to explore new physical mechanisms capable of reconciling these once promising models with the updated data \cite{Kallosh:2025act,Aoki:2025,Berera:2025,Dioguardi:2025,Salvio:2025,Antoniadis:2025,Kim:2025,Drees:2025,Haque:2025,Peng:2025bws,Liu:2025qca,Gialamas:2025ofz,Wolf:2025ecy,Dioguardi:2025mpp,Heidarian:2025drk,
Raidal:2025,Vagnozzi}. For instance, Kallosh et al. \cite{Kallosh:2025act} investigated modifications to the inflationary potentials, demonstrating that slight adjustments in the functional form can reconcile classic models with the latest observational constraints. In a different approach, Berera et al. \cite{Berera:2025} highlighted the role of warm inflation, arguing that dissipative effects naturally shift the scalar spectral index toward larger values without requiring drastic changes to the potential. Furthermore, Gialamas et al. \cite{Gialamas:2025ofz} analyzed models based on modified gravity, showing that additional degrees of freedom can effectively alleviate the tension in Starobinsky like models, thereby extending their observational viability.

On the other hand, the increasing best fit value of $n_s$ creates a new opportunity for other inflationary models. Potentials that naturally predict a large $n_s$ now have the chance to fall into the favored region of the recent joint data. A notable example is the family of fractional power law potentials $V(\varphi) \propto \varphi^n$. Earlier observations had heavily ruled out these potentials. However, currently, their predictions for $n \leq 2/3$ can successfully fall within the 95\% CL of the latest data.
These developments strongly motivate us to search for physical mechanisms that could further enhance the consistency of these potentials with the recent joint datasets (P-ACT-LB-BK18) \cite{Planck:2018inflation1, bk18, ACT:DR6params, ACT:DR6models}, aiming to bring their predictions well within the 68\% CL.

Among the various approaches pursued in this direction, Loop Quantum Cosmology (LQC) has emerged as one of the most compelling frameworks for quantizing space time and resolving the singularity problem.  By presenting a picture of a Big Bounce instead of the Big Bang, LQC makes the history of the universe traceable to before the Big Bang singularity \cite{Barrau:2014,Ashtekar1:2006,Ashtekar2:2008,Abolhasan1:2024}. Owing to the discrete nature of space time at the Planck scale, LQC predicts the existence of certain quantum corrections to the classical equations of motion. Two main types of corrections are of particular importance: holonomy and inverse volume corrections
\cite{Barrau:2014,Ashtekar1:2006,Ashtekar2:2008,Abolhasan1:2024,Bojowald0:2011,Bojowald:2011,Calcagni:2011,Zhu:2015,Zhu:2016,parvizi:2026JHEAP1,Mielszarek1:2008,Mielszarek2:2008}.

In the holonomy approach, the modified Friedmann and Klein-Gordon equations in the Friedmann–Lema\^{\i}tre–Robertson–Walker (FLRW) background not only prevent the Big Bang singularity but also explicitly account for scalar, vector, and tensor perturbations. However, it has been shown that the magnitude of these corrections in the perturbation equations is extremely small, on the order of $\mathcal{O}(10^{-12}$), implying that extremely high precision is required to detect or compute these effects \cite{Barrau:2014}.

On the other hand, inverse volume corrections become significant at very high densities due to the existence of a minimal measurable volume in nature. These corrections introduce modifications to the background Friedmann and Klein-Gordon equations and, at the perturbation level, alter the evolution of the scalar and tensor perturbations. As a result, inverse volume corrections can have a more pronounced influence on a broad range of observable predictions, including the power spectrum of density perturbations, the scalar spectral index, and the tensor-to-scalar ratio \cite{Barrau:2014,Bojowald:2011,Calcagni:2011,Zhu:2015,Zhu:2016}.

In this study, we aim to examine the consistency of inflationary models based on the fractional power law potentials within the framework of LQC with inverse volume corrections, against current observational evidence. Specifically, we seek to determine whether LQC predictions concerning modifications to the primordial power spectrum can improve the compatibility between these models and recent ACT observational data.

Building upon our previous investigation into LQC inverse volume corrections \cite{parvizi:2026JHEAP1}, the present study adopts a more refined analytical framework. While foundational LQC analyses - including our prior work - have predominantly relied on leading order slow roll approximations to evaluate the perturbation equations \cite{Bojowald:2011,Calcagni:2011}, we now incorporate the second order slow roll formalism developed by Zhu et al. \cite{Zhu:2015, Zhu:2016}. This higher order precision provides a more rigorous evaluation of the dynamical phase space, enabling a highly reliable assessment of how inverse volume corrections impact the theoretical predictions of various scalar potentials. Consistent with standard cosmological perturbation theory, all inflationary observables are systematically expanded and evaluated at the epoch of Hubble horizon crossing \cite{Zhu:2016}.

While the impact of LQC inverse volume corrections on the background evolution is subdominant during the slow roll phase , these modifications leave significant and potentially measurable imprints on cosmological observables, particularly the spectral indices and their runnings. Consequently, this study adopts a semi-classical effective framework, providing a consistent theoretical basis to rigorously test these quantum gravitational consequences against high precision observational data.

The paper is organized as follows. In Section \ref{sec2}, we briefly review the theoretical foundations of LQC and introduce the mathematical formulation of the inverse volume corrections. In Section \ref{sec:observational_viability}, we analyze the fractional power law potentials for the specific cases of $n=(1/3, 2/5, 2/3)$. By incorporating the modified perturbation equations, we systematically confront the theoretical predictions of these models with the latest joint observational data. Finally, Section \ref{sec:conclusion} is devoted to a discussion of our results and concluding remarks.

\section{Background equation and perturbations in LQC}\label{sec2}
In the standard inflationary model of cosmology, the action is given by \cite{Guth:1981,Linde:1982}
\begin{equation}\label{eq:action}
S=\int{\rm d}^{4}x \sqrt{-g} \left[\frac{M_p^2}{2}R+ X - V(\varphi) \right] ,
\end{equation}
where  $M_p \equiv 1 / {\sqrt{8\pi G}}$ indicates the reduced Planck mass, $g$ is the determinant of the metric tensor $g_{\mu\nu}$ and $R$ is the Ricci scalar.
The kinetic energy term and the scalar field potential are denoted by $X \equiv \frac{1}{2}g^{\mu\nu}\partial_\mu \varphi \partial_\nu \varphi$ and $V(\varphi)$, respectively.

In the framework of LQC, the quantization of space time geometry introduces specific corrections that modify the classical dynamics.
Here, we consider a spatially flat FLRW universe described by the conformal time metric ${\rm d}s^2 = a^2(\tau)(-{\rm d}\tau^2 + {\rm d}x^i {\rm d}x_i)$, where $a(\tau)$ represents the scale factor. The LQC effective Friedmann and Klein-Gordon equations then take the following forms \cite{Bojowald:2011}
\begin{align}
	&\mathcal{H}^2 = \frac{8\pi G}{3} \alpha\left[\frac{1}{2\nu}
	(\varphi')^2 + pV(\varphi) \right], \label{eq:FR1-ML}\\
	&	\varphi'' + 2\mathcal{H}\left(1-\frac{{\rm d}\ln
		\nu}{{\rm d}\ln p}
	\right)\varphi' +
	\nu pV_{,\varphi} = 0, \label{eq:KL-GO}
\end{align}
where $\mathcal{H} \equiv \frac{a'}{a}$ is the conformal Hubble parameter and $p\equiv a^2$.
Additionally, the prime $'$ and the subscript $,\varphi$ indicate the derivatives with respect to the conformal time $\tau$ and the scalar field $\varphi$, respectively.

Furthermore, the correction functions $\alpha$ and $\nu$ arise from inverse volume quantum geometry effects inherent to LQC, reflecting the underlying discrete structure of spacetime. Within an effective semiclassical framework, these corrections are derived via an effective expansion for a homogeneous and isotropic FLRW background, following the methodologies established in \cite{Barrau:2014,Bojowald:2011,Calcagni:2011}. These functions encode the deviations of the inverse volume operators from their classical limits, deviations that are notably sensitive to the specific lattice refinement scheme. In this work, we adopt a standard effective expansion in which $\alpha$ and $\nu$ correspond to the leading order contributions, thereby encapsulating the dominant impact of quantum geometry on the cosmological dynamics and take the following forms 
\begin{equation}\label{eq:alpha-LQC}
	\alpha \simeq 1 + \alpha_0 \delta_{\rm pl},
\end{equation}
\begin{equation}\label{eq:nu-LQC}
	\nu \simeq 1 + \nu_0 \delta_{\rm pl}.
\end{equation}
Here, $\delta_{\rm pl} \equiv (a_{\rm pl}/a)^\sigma$ represents the evolving quantum correction, which changes with the scale factor $a$. The parameters $\sigma$, $\alpha_0$, $\nu_0$, and $a_{\rm pl}$ are constants that can be fixed by the specific loop quantization scheme. Specifically, the parameter $\sigma$ represents the power law index governing the scaling behavior of the inverse volume corrections with respect to the scale factor. Physically, $\sigma$ characterizes the rate at which the effects of spacetime discreteness diminish during cosmic expansion. This parameter is deeply rooted in the distribution of the background quantum state within the spin network framework of LQG, thereby providing a quantitative measure of the underlying geometric granularity. Furthermore, along with $\alpha_0$ and $\nu_0$, the parameter $\sigma$ acts as a phenomenological parameter that allows the effective model to be constrained by and reconciled with cosmological observations. Note that if we turn off the inverse volume corrections by setting $\alpha=\nu=1$, Eqs. (\ref{eq:FR1-ML})-(\ref{eq:KL-GO}) simply reduce to the standard background equations for inflation.

The effective equations are evaluated in the semi-classical regime and remain valid only up to linear order in the quantum correction parameter $\delta_{\rm pl}$. For consistency, all subsequent calculations expand quantities strictly to the first order in $\delta_{\rm pl}$.
While these inverse volume corrections introduce purely geometric modifications to the Friedmann and Klein-Gordon equations, their impact on the background trajectory during the slow roll phase is strictly subdominant. This ensures that the background dynamics remain largely consistent with standard slow roll inflation.

To achieve a framework with higher predictability, the number of independent parameters can be reduced by imposing a consistency condition derived from the requirements of an anomaly free constraint algebra.
For $\sigma \neq 3$, one can estimate the relation between $\nu_0$ and $\alpha_0$ as follows
\begin{equation} \label{eq:alfa0_nu0_in_LQC}
	\nu_0 = \dfrac{3 (\sigma-6)}{(\sigma+6)(\sigma-3)} \alpha_0.
\end{equation}
It is noteworthy that while the parameter $\sigma$ can vary in the range $[0,6]$, smaller values are theoretically favored. Furthermore, It has been shown that the observable effects of quantum gravity become undetectable for $\sigma \geq 2$ \cite{Bojowald:2011,Calcagni:2011}. Therefore, to ensure a comprehensive analysis that fully captures the detectable region and the transition where these quantum effects fade away, we restrict our study to the parameter range $\sigma \in [0,3] $.

Since the aim of studying slow roll inflation with quantum corrections is to utilize a mathematically consistent framework, one must restrict the analysis to a region where the effective model is anomaly free. In quantum cosmology, quantum effects dominate in the highly compressed, ultra dense regimes near the Big Bang singularity, which LQC aims to resolve. However, in the deep quantum regime, incorporating inverse volume corrections often leads to a failure in the closure of the constraint algebra \cite{Bojowald0:2011,Calcagni:2011}.
This generates anomalies that break general covariance, meaning an effective space time description fails. Consequently, to study slow roll inflation robustly, one must evaluate the theory in the semi-classical or large volume limit, where the constraint algebra can be closed and equations remain well behaved, even though the quantum corrections ($\delta_{\rm pl}$) are small \cite{Zhu:2016}.

To solve the dynamics analytically, one can employ a double expansion approximation method \cite{Zhu:2016}. In this approach, the quantum parameter ($\delta_{\rm pl} \ll 1$) is treated perturbatively, while simultaneously utilizing the standard inflationary slow roll parameters ($\epsilon_i$) to navigate this stable regime. The $\epsilon_i$ parameters measure the fractional change of the Hubble rate and are defined hierarchically as
\begin{equation}
	\epsilon_1 \equiv - \frac{{\rm d} \ln H}{{\rm d} \ln a}, \hspace{1em}\hspace{1em}
	\epsilon_{i+1} \equiv \frac{{\rm d} \ln\epsilon_i}{{\rm d}\ln a}. \label{eq:epsilon_i}
\end{equation}
By applying the uniform asymptotic approximation method, the power spectra for both scalar and tensor perturbations evaluated at horizon crossing can then be analytically derived as follows \cite{Zhu:2016}
\begin{align}
	\mathcal{P} _s(k) &\approx A_S^{\star}
	\Bigg[
	1 - 0.5431\epsilon_{\star1} +0.7284\epsilon_{\star2}
	-0.4580\epsilon_{\star1}^2
	+ 0.4997\epsilon_{\star2}^2
	+ 0.0505\epsilon_{\star1}\epsilon_{\star2} \nonumber \\
	&+ 0.1616\epsilon_{\star2}\epsilon_{\star3}
	+ \epsilon_{\rm pl}\left(\frac{3H_{\star}}{2}\right)^{\sigma}
	\left(
	\frac{Q_{-1}^{\star (s)}}{\epsilon_{\star1}} + Q_{0}^{\star (s)} + \frac{Q_{1}^{\star (s)}\epsilon_{\star2}}{\epsilon_{\star1}}
	\right)
	\Bigg],\label{eq:power_spectrum_s1}
\end{align}
where $A_S^{\star}\equiv\frac{181H_{\star}^2}{72e^3\pi^2\epsilon_{\star1}} $, $\epsilon_{\rm pl}\equiv (a_{\rm pl}/k)^{\sigma}$ and the subscript $"\star"$ denotes evaluation at horizon crossing. Also the quantities $Q_{-1}^{\star(s)} $, $	Q_{0}^{\star(s)} $ and $	Q_{1}^{\star(s)} $ are defined in Appendix A. 
 Planck measurements have constrained the scalar power spectrum to ${\mathcal{P} _s(k_\ast) \simeq 2.1 \times 10^{-9}}$ at the CMB pivot scale $k_\ast = 0.05~{\rm Mpc^{-1}}$ \cite{Planck:2018inflation,bk18}.

Similarly, the tensor power spectrum can be obtained as follows \cite{Zhu:2016} 
\begin{align}
	\mathcal{P}_t(k) &\approx A_t^{\star}
	\Bigg[
	1 - 0.5431\epsilon_{\star 1} -0.4580\epsilon_{\star 1}^2 \nonumber \\
	&\quad -0.2199\epsilon_{\star 1}\epsilon_{\star 2} + \epsilon_{\text{pl}}\left(\frac{3H_{\star}}{2}\right)^{\sigma}Q_{0}^{\star(t)}
	\Bigg],	\label{eq:power_spectrum_t}  	
\end{align}
where $A_t^{\star}$ is the amplitude of tensor perturbations and 
\begin{equation}
Q_{0}^{\star(t)}=-\frac{2\sigma(\sigma^2-2\sigma+6)}{3}\alpha_0L_1-\frac{80\sigma(\sigma-3)}{543}\alpha_0L_3
+\frac{240}{181}\alpha_0L_4,
\end{equation}
in which the quantities $L_1 $, $L_3$ and $L_4$ are defined in Appendix A.
The scalar spectral index $n_{\rm s}$ and the tensor-to-scalar ratio $r$ are defined as
\begin{equation}
		n_{\rm s} - 1 \equiv \frac{{\rm d}\ln \mathcal{P}_s(k)}{{\rm d}\ln k}, \label{eq:ns_in_LQC}
	\end{equation}
	\begin{equation}
		r\equiv \dfrac{	\mathcal{P}_t(k)}{\mathcal{P}_s(k)}. \label{eq:r_in_LQC}
	\end{equation}
By substituting the power spectra from Eqs. (\ref{eq:power_spectrum_s1}) and (\ref{eq:power_spectrum_t}) into these definitions, one can evaluate $n_{\rm s}$ and $r$ analytically. Up to the second order in the slow roll approximation, the results are as follows
\begin{align}
	n_{\rm s} &\approx 1 - 6\epsilon_V + 2\eta_V +14.1483 \epsilon_V^2-13.6544 \epsilon_V \eta_V+\frac{2}{3}\eta_V^2\nonumber \\ &+2.1235 \xi_V^2 +\frac{\epsilon_{\rm pl}H_\star ^\sigma}{\epsilon_V}\left\lbrace \frac{3^\sigma}{2^\sigma} K_{-1}^{\star(s)}+
	\frac{\sigma^2 (\sigma-3)\alpha_0}{18}(-3.1852 \sigma-3)\right\rbrace ,  \label{eq:ns1_in_LQC}
\end{align}
\begin{align}
	r &\approx 16\epsilon_V\Bigg\{ 1-4.2469\epsilon_V+2.1235\eta_V\nonumber \\
	&\qquad+\frac{\epsilon_{\rm pl}H_\star ^\sigma}{\epsilon_V}
	\left[ \frac{3^\sigma}{2^\sigma} \frac{K_{-1}^{\star(s)}}{\sigma}+\frac{\sigma(\sigma^2-9)(-0.7284 \sigma-1)\alpha_0}{18}\right]\Bigg\},\label{eq:r1_in_LQC}
\end{align}
where the quantities $\epsilon_{\rm V} \equiv \frac{M_p^2}{2} \left( \frac{V_{,\varphi}}{V} \right)^2$, $\eta_{\rm V} \equiv M_p^2 \left(\frac{V_{,\varphi\varphi}}{V}\right)$, and $\xi_{\rm V} \equiv M_p^4 \left(\frac{V_{,\varphi} V_{,\varphi\varphi\varphi}}{V^2}\right)$ denote the standard potential slow roll parameters. 

Note that Eqs. (\ref{eq:ns1_in_LQC}) and (\ref{eq:r1_in_LQC}) are valid for $\sigma\neq 3$. For the case of $\sigma=3$ we have $\alpha_0=0$ and one must replace $K_{-1}^{\star(s)} $ by $K_{0}^{\star(s)} \epsilon_{\star1}$ in Eqs. (\ref{eq:ns1_in_LQC}) and (\ref{eq:r1_in_LQC}) \cite{Zhu:2015}. Here
	\begin{equation}
		K_{0}^{\star(s)}=-1.5432 \sigma Q_{-1}^{\star(s)}- \sigma Q_{0}^{\star(s)} ,\label{eq:K0star-s}
	\end{equation}
	in which the quantities $Q_{-1}^{\star(s)} $ and $Q_{0}^{\star(s)}$ given by appendix A are evaluated at $\sigma=3$ and $\alpha_0=0$. Also from \cite{Calcagni:2011} for $\sigma=3$ since $\alpha_0=0$ we obtain $\epsilon_{\star 1} \approx\left(1-\frac{1}{2}\nu_0\delta_{\rm pl*}\right)\epsilon_{V}=\big(1-\delta(k_*)/2\big)\epsilon_{V}$.

The latest observational constraints on the scalar spectral index have been derived from the joint analysis of ACT DR6, Planck 2018, DESI BAO, and BICEP/Keck datasets, yielding a value of $n_s = 0.974 \pm 0.003$~\cite{ACT:DR6params, ACT:DR6models}. Furthermore, recent measurements from Planck and BICEP/Keck 2018 place an upper bound on the tensor-to-scalar ratio at $r < 0.036$~\cite{bk18}.
		
In the following section, we apply the LQC framework to evaluate specific fractional power law potentials against the aforementioned high precision datasets. To ensure the mathematical validity of the linear approximations employed in our effective equations, we must strictly impose the condition $\alpha_0\delta_{\rm pl} < 1$. Operating within this valid perturbative regime, and restricting our parameter space to $\sigma \in [0, 3]$ as previously justified, we will investigate whether these quantum gravity corrections can successfully reconcile the theoretical predictions for $n_{\rm s}$ and $r$ with the latest cosmological observations.
\section{Fractional power law potentials in LQC}\label{sec:observational_viability}
In this section, we investigate the observational viability of fractional power law inflationary potentials within the context of LQC inverse volume corrections. The general form of this potential is given by \cite{Planck:2018inflation1}
  \begin{equation}\label{eq:power_law_pot}
  	V(\varphi) = V_0 \varphi^n,
  \end{equation}
where $n=(1/3,2/5,2/3)$ and the parameter $V_0$ is constant and is fixed by the value of scalar power spectrum at the CMB pivot scale $k_* = 0.05~{\rm Mpc^{-1}}$.
Consistent with our semi-classical approach, we assume that the background evolution of the scalar field $\varphi$ and the conformal Hubble parameter $\mathcal{H}$ remains purely classical. However, the quantum gravitational effects arising from LQC inverse volume corrections are rigorously incorporated into the dynamics of the cosmological perturbations.

For the potential (\ref{eq:power_law_pot}), the potential slow roll parameters can be calculated as follows
\begin{equation}
	\epsilon_{\rm V}(\varphi) = \frac{n^2 M_p^2}{2 \varphi^2}, \label{eq:epsilon_v1}
\end{equation}
\begin{equation}
	\eta_{\rm V}(\varphi) =\frac{ n (n-1) M_p^2}{ \varphi^2}= \frac{2(n-1)}{n} \epsilon_V(\varphi), \label{eq:eta_v1}
\end{equation}
\begin{equation}
		\xi_{\rm V}(\varphi)= \frac{ n^2(n-1) (n-2) M_p^4}{ \varphi^4}= \frac{ 4(n-1)(n-2)}{ n^2}\epsilon_{\rm V}^2(\varphi). \label{eq:kesay_v1}
\end{equation}
To evaluate the observables at horizon crossing, we must determine the field value $\varphi$ as a function of $e$-folds number $N$. In the slow roll approximation, this background evolution is governed by the differential equation as follows
\begin{equation}
	\varphi' = \dfrac{a^2 V'(\varphi)}{3 \mathcal{H}^2}.
\end{equation}
The integration constant for this equation is fixed by determining the field value at the end of inflation, $\varphi_{\rm end}$, which is established by the condition $\epsilon_V(\varphi_{\rm end}) = 1$.
Furthermore, to parameterize the LQC inverse volume quantum corrections at the CMB pivot scale $k_*$, when $\sigma \neq 3$ we use the following relation
\begin{equation} \label{eq:delta(k)1_in_LQC}
	\delta(k_*) =\alpha_0 \epsilon_{\rm pl*}H_{\star}^{\sigma}=\alpha_0\delta_{\rm pl*},
\end{equation}
where $\epsilon_{\rm pl*}\equiv(a_{\rm pl}/k_*)^{\sigma}\ll1.$ For $\sigma=3$ one has $\alpha_0=0 $  \cite{Calcagni:2011}, in which case Eq. (\ref{eq:delta(k)1_in_LQC}) is replaced by
	\begin{equation} \label{eq:delta(k)2_in_LQC}
		\delta(k_*) =\nu_0 \epsilon_{\rm pl*}H_{\star}^{\sigma}=\nu_0\delta_{\rm pl*}.
\end{equation} 

Consequently, the influence of these quantum geometric effects on the primordial perturbations can be fundamentally encapsulated by the power exponent $\sigma$ and the effective amplitude $\delta(k_*)$ at the CMB pivot scale.
%
%
%
%
%
%

\begin{figure}[H]
	\centering
	
	\begin{subfigure}{0.47\textwidth}
		\centering
		\includegraphics[width=\textwidth]{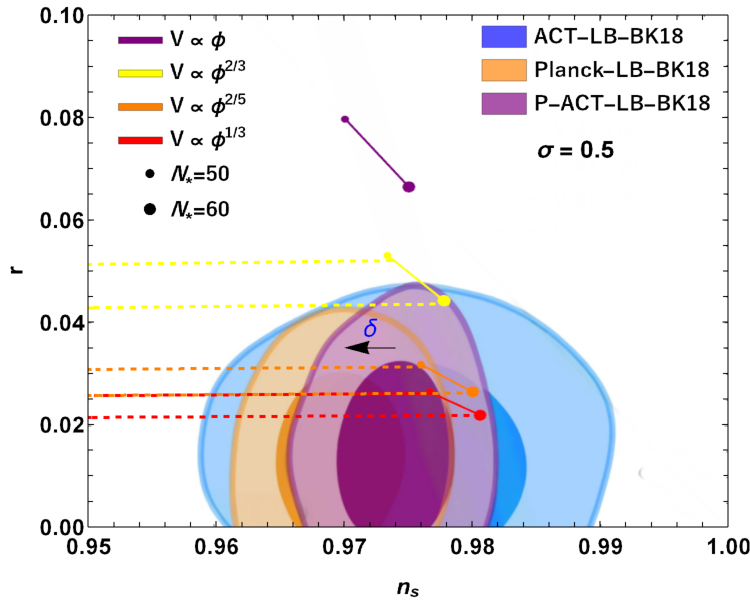}
		\label{fig-power-law_LQC1a}
		\caption{}
	\end{subfigure}
	\hfill
	\begin{subfigure}{0.47\textwidth}
		\centering
		\includegraphics[width=\textwidth]{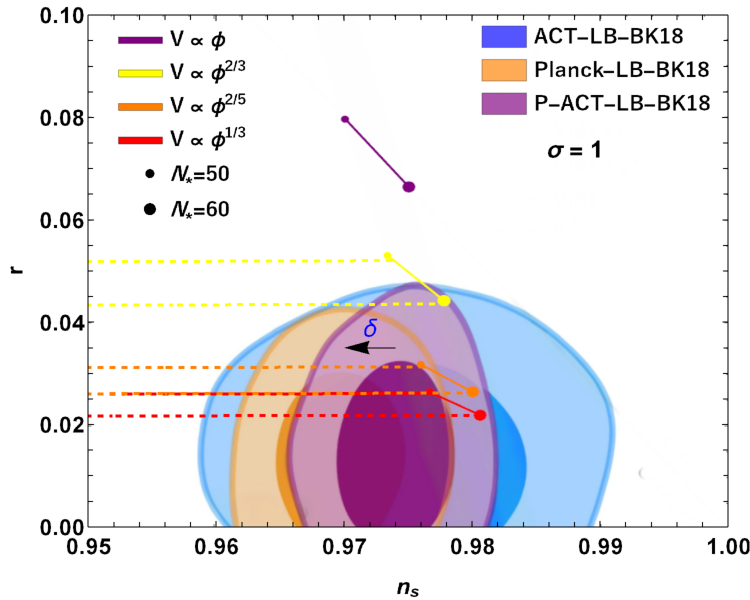}
		\label{fig-power-law_LQC1b}
		\caption{}
	\end{subfigure}
	
	\medskip
	
	\begin{subfigure}{0.47\textwidth}
		\centering
		\includegraphics[width=\textwidth]{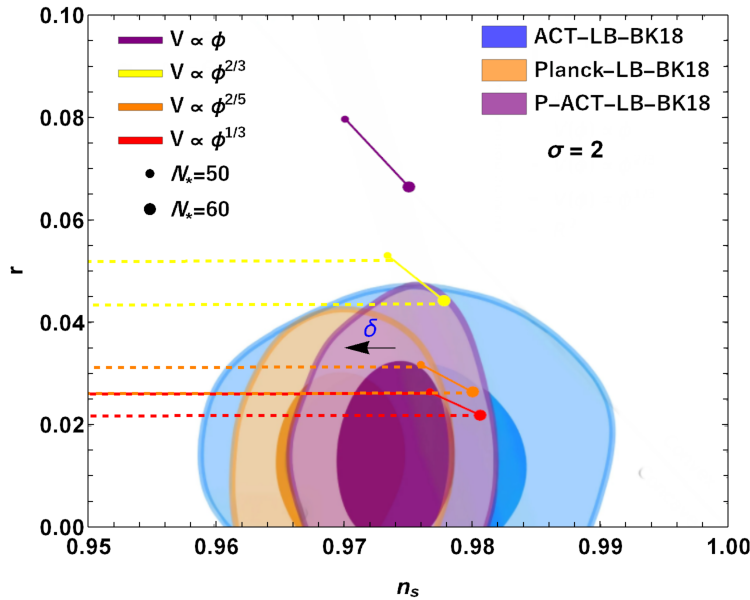}
		\label{fig-power-law_LQC1c}
		\caption{}
	\end{subfigure}
	\hfill
	\begin{subfigure}{0.47\textwidth}
		\centering
		\includegraphics[width=\textwidth]{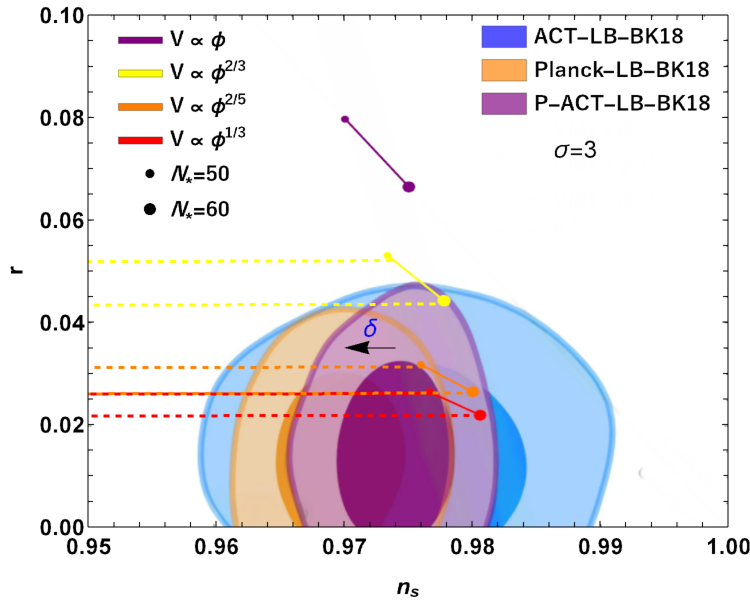}
		\label{fig-power-law_LQC1d}
		\caption{}
	\end{subfigure}
	
	\caption{Predictions for the tensor-to-scalar ratio $r$ versus the scalar spectral index $n_{\rm s}$ for the fractional power law potentials $V(\varphi) \propto \varphi^n$ $(n = 1/3, 2/5, 2/3)$ incorporating LQC inverse volume corrections. The dashed curves represent the evolutionary trajectories between $N=50$ and $N=60$ $e$-folds. In all panels, the amplitude parameter $\delta$ is varied over the range $[0, 10^{-2}]$, while the exponent $\sigma$ is fixed at $0.5, 1, 2,$ and $3$ for panels (a), (b), (c), and (d), respectively. The black arrows indicate the direction in which the varying parameter $\delta$ increases.}
	\label{fig-power-law_LQC1}
\end{figure}

In the absence of LQC corrections, where $\delta=0$, the predictions of the standard inflationary scenario for the fractional power law potentials $V(\varphi) \propto \varphi^n$ with $n=(1/3, 2/5, 2/3)$ are represented in Fig. \ref{fig-power-law_LQC1} by the solid points in the $r-n_{\rm s}$ plane, bounded by the $e$-foldings $N=50$ and $N=60$. According to Fig. \ref{fig-power-law_LQC1}, confronting these classical baselines with the joint P-ACT-LB-BK18 constraints reveals a distinct hierarchy of viability. For the case of $n=2/3$, the $N=50$ prediction is similarly ruled out, while the $N=60$ prediction marginally enters the 95\% CL, resting close to the boundary. The case $n=2/5$ exhibits improved viability, with both $e$-folding values falling within the 95\% CL; notably, the $N=50$ prediction just touches the boundary of the tighter 68\% CL region.Finally, the case $n=1/3$ is the most favored in the classical regime, with its $N=60$ prediction situated within the 95\% CL and the $N=50$ prediction marginally entering the 68\% CL.

The focus on fractional potentials is motivated by the fact that LQC inverse volume corrections, modulated by $\sigma$ and $\delta$, tend to shift the predictions of these models in the $r-n_{\rm s}$ plane, effectively pulling fractional exponent models into the observational confidence regions where integer power law models (e.g., $n=1, 2, 3, 4$) remain excluded. 

\begin{figure}[H]
	\centering
	
	\begin{subfigure}{0.47\textwidth}
		\centering
		\includegraphics[width=\textwidth]{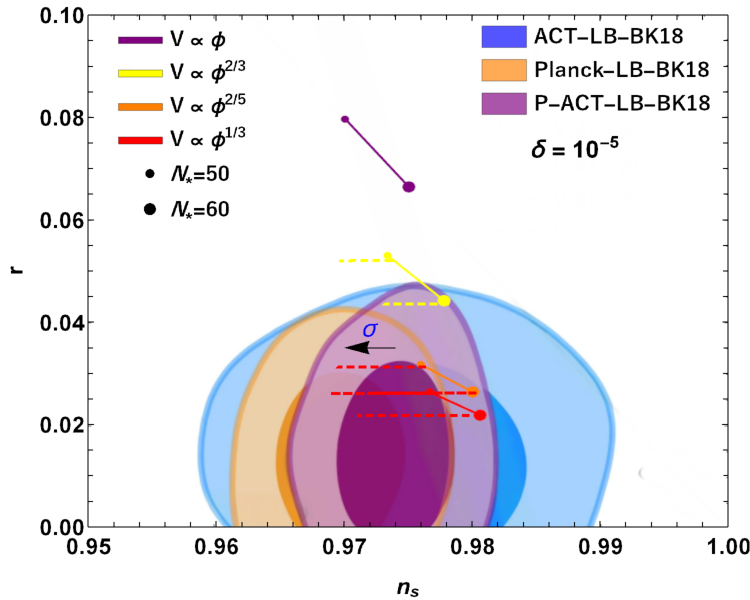}
		\label{fig-power-law_LQC2d}
		\caption{}
	\end{subfigure}
	\hfill
	\begin{subfigure}{0.47\textwidth}
		\centering
		\includegraphics[width=\textwidth]{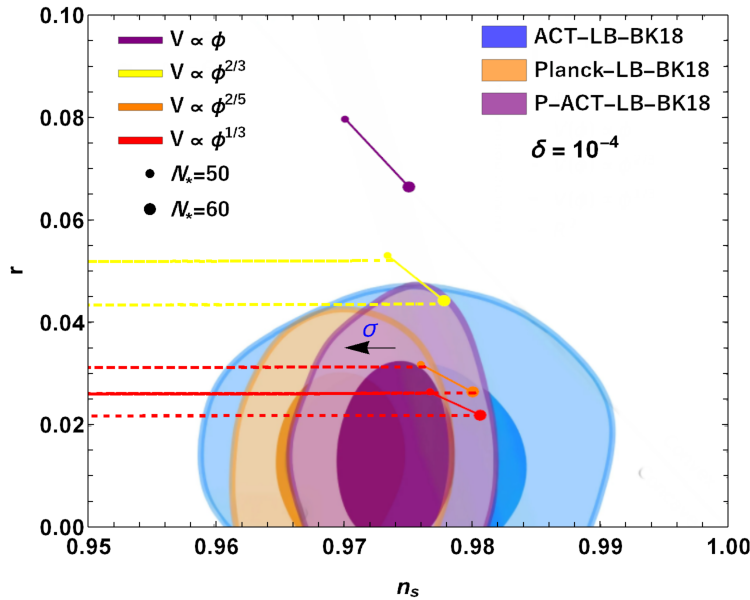}
		\label{fig-power-law_LQC2c}
		\caption{}
	\end{subfigure}
	
	\medskip
	
	\begin{subfigure}{0.47\textwidth}
		\centering
		\includegraphics[width=\textwidth]{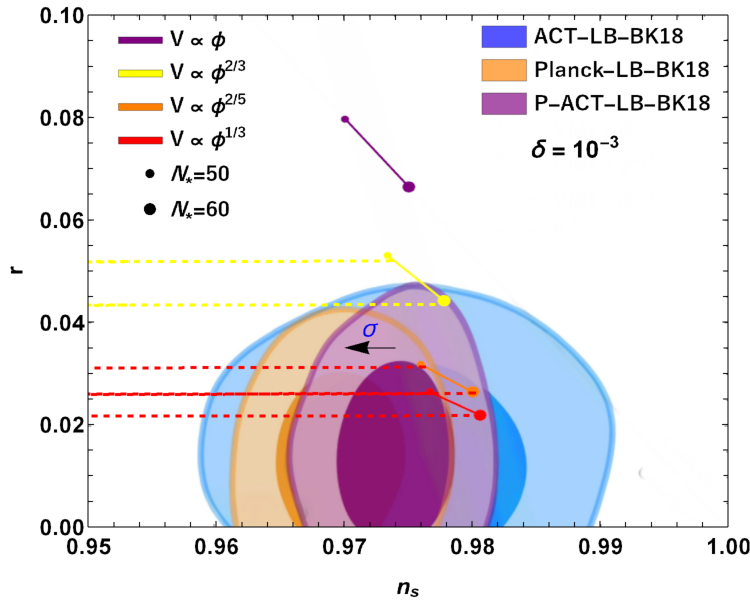}
		\label{fig-power-law_LQC2b}
		\caption{}
	\end{subfigure}
	\hfill
	\begin{subfigure}{0.47\textwidth}
		\centering
		\includegraphics[width=\textwidth]{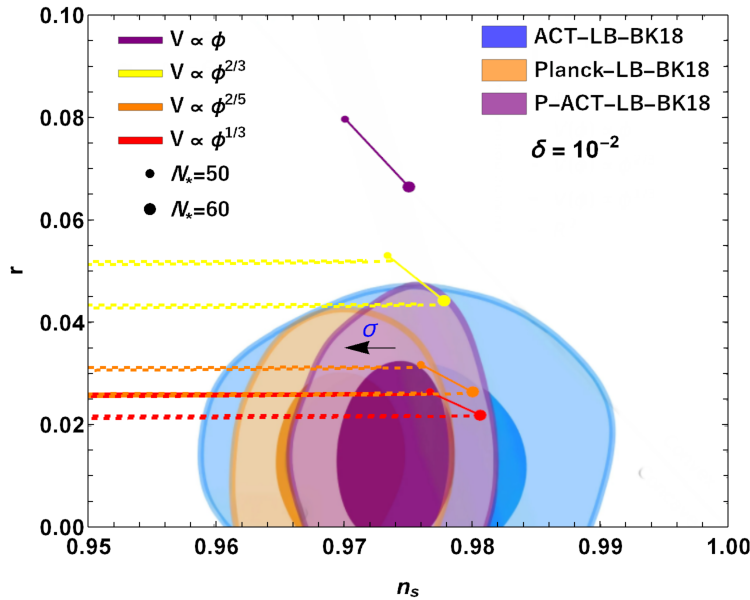}
		\label{fig-power-law_LQC2a}
		\caption{}
	\end{subfigure}
	
	\caption{Same as Fig. \ref{fig-power-law_LQC1}, but the exponent $\sigma$ is varied over the range $[0, 3]$, while the amplitude parameter $\delta$ is fixed at $10^{-5}, 10^{-4}, 10^{-3},$ and $10^{-2}$ for panels (a), (b), (c), and (d), respectively. The black arrows indicate the direction in which the varying parameter $\sigma$ increases.}
	\label{fig-power-law_LQC2}
\end{figure}

As demonstrated in Figs. \ref{fig-power-law_LQC1} and \ref{fig-power-law_LQC2}, the inclusion of these corrections induces a significant horizontal shift of the predicted values to the left. This effect is primarily introduced through the additional terms in the expression for $n_{\rm s}$ in Eq. (\ref{eq:ns1_in_LQC}) and is governed by the LQC parameters $\delta$ and $\sigma$, allowing the model predictions to be steered into observationally favored regions (the 68\% and 95\% CLs).
 Analysis confirms that the primary effect is the modification of the scalar spectral index, $n_{\rm s}$, while the contribution to the tensor-to-scalar ratio, $r$, remains sub-dominant.
This is consistent with the theoretical framework, as the correction term in Eq. (\ref{eq:ns1_in_LQC}) is more significant than that in Eq. (\ref{eq:r1_in_LQC}). Consequently, increasing either $\delta$ (for a constant $\sigma$ in Fig. \ref{fig-power-law_LQC1}) or $\sigma$ (for a constant $\delta$ in Fig. \ref{fig-power-law_LQC2}) translates the model predictions almost horizontally from right to left across the $r-n_{\rm s}$ plane.

%
%
%
To fully illustrate these effects across the fractional power law potentials, we present the resulting $r-n_{\rm s}$ diagrams in two configurations.
Figure \ref{fig-power-law_LQC1} corresponds to the case where the parameter $\delta$ is varied over the range $[0,10^{-2}]$ while $\sigma$ is fixed at $0.5, 1, 2,$ and $3$. Conversely, Figure \ref{fig-power-law_LQC2} corresponds to the case where the parameter $\sigma$ is varied over the range $[0,3]$ while $\delta$ is fixed at $10^{-5}, 10^{-4}, 10^{-3},$ and $10^{-2}$. In these figures, the diagrams obtained by systematically varying $\delta$ while holding $\sigma$ constant, or vice versa, exhibit similar structural behavior, thus confirming the robust nature of the horizontal translation mechanism described previously.

While the graphical variations with respect to $\sigma$ and $\delta$, respectively, in Figs.~\ref{fig-power-law_LQC1} and \ref{fig-power-law_LQC2} are subtle, the corresponding quantitative dependencies are explicitly captured in Tables~\ref{tab:merged_delta} and \ref{tab:merged_sigma}. To systematically quantify these observationally viable regions, the allowed ranges for the LQC inverse volume parameters are consolidated in these tables. Table~\ref{tab:merged_delta} details the constraints on the amplitude $\delta$ for fixed values of the exponent $\sigma$, while Table~\ref{tab:merged_sigma} provides the reciprocal constraints on $\sigma$ for fixed $\delta$. By examining the data across the fractional power law potentials with $n=(1/3, 2/5, 2/3)$, several key physical trends become apparent. Most notably, as the potential becomes steeper (larger $n$), the observationally permitted parameter space shrinks significantly. For instance, the $n=2/3$ potential is entirely ruled out at $N=50$, rendering that portion of the parameter space empty, and is only marginally viable at $N=60$. In contrast, the shallower $n=1/3$ potential accommodates a much broader range of quantum geometric corrections.

\begin{table}[h!]
	\centering
	\caption{Allowed ranges for the LQC inverse volume parameter $\delta$ for different values of $\sigma$, based on 68\% and 95\% CLs constraints from P-ACT-LB-BK18 data. Results are compared across the fractional power law potentials with $n=(1/3, 2/5, 2/3)$. The column for $n=2/3$ at $N=50$ is omitted because it lies entirely outside the observational bounds. The overall order of magnitude for each row is provided in the rightmost factor column.}
	\label{tab:merged_delta}
	\resizebox{\textwidth}{!}{
		\begin{tabular}{c | c c | c c | c c | c c | c c | c}
			\hline\hline
			& \multicolumn{4}{c|}{$n = 1/3$} & \multicolumn{4}{c|}{$n = 2/5$} & \multicolumn{2}{c|}{$n = 2/3$} & \\
			\cline{2-11}
			& \multicolumn{2}{c|}{$\delta$ ($N=50$)} & \multicolumn{2}{c|}{$\delta$ ($N=60$)} & \multicolumn{2}{c|}{$\delta$ ($N=50$)} & \multicolumn{2}{c|}{$\delta$ ($N=60$)} & \multicolumn{2}{c|}{$\delta$ ($N=60$)} & Factor \\
			\cline{2-11}
			$\sigma$ & 95\% CL & 68\% CL & 95\% CL & 68\% CL & 95\% CL & 68\% CL & 95\% CL & 68\% CL & 95\% CL & 68\% CL & \\
			\hline
			$0.5$ & $[0, 1.95]$ & $[0, 1.22]$ & $[0, 2.40]$ & $[0.41, 1.73]$ & $[0, 2.06]$ & $[0.06, 0.87]$ & $[0, 2.60]$ & $[0.48, 1.82]$ & $[0, 1.98]$ & - & $\times 10^{-3}$ \\
			$1$ & $[0, 1.64]$ & $[0, 1.04]$ & $[0, 2.12]$ & $[0.38, 1.58]$ & $[0, 1.74]$ & $[0.05, 0.77]$ & $[0, 2.30]$ & $[0.43, 1.58]$ & $[0, 1.74]$ & - & $\times 10^{-4}$ \\
			$2$ & $[0, 1.87]$ & $[0, 1.13]$ & $[0, 2.25]$ & $[0.41, 1.67]$ & $[0, 1.92]$ & $[0.06, 0.81]$ & $[0, 2.50]$ & $[0.48, 1.73]$ & $[0, 1.83]$ & - & $\times 10^{-5}$ \\
			$3$ & $[0, 14.80]$ & $[0, 9.10]$ & $[0, 21.30]$ & $[4.12, 15.90]$ & $[0, 12.70]$ & $[0.70, 5.10]$ & $[0, 19.70]$ & $[4.10, 14.0]$ & $[0, 8.20]$ & - & $\times 10^{-4}$ \\
			\hline\hline
		\end{tabular}
	}
\end{table}

\begin{table}[h!]
	\centering
	\caption{Allowed ranges for the LQC inverse volume parameter $\sigma$ for different values of $\delta$, based on 68\% and 95\% CLs constraints from P-ACT-LB-BK18 data. Results are compared across the power law potentials $n=(1/3, 2/5, 2/3)$. The column for $n=2/3$ at $N=50$ is omitted because it lies entirely outside the observational bounds.}
	\label{tab:merged_sigma}
	\resizebox{\textwidth}{!}{
		\begin{tabular}{c | c c | c c | c c | c c | c c}
			\hline\hline
			& \multicolumn{4}{c|}{$n = 1/3$} & \multicolumn{4}{c|}{$n = 2/5$} & \multicolumn{2}{c}{$n = 2/3$} \\
			\cline{2-11}
			& \multicolumn{2}{c|}{$\sigma$ ($N=50$)} & \multicolumn{2}{c|}{$\sigma$ ($N=60$)} & \multicolumn{2}{c|}{$\sigma$ ($N=50$)} & \multicolumn{2}{c|}{$\sigma$ ($N=60$)} & \multicolumn{2}{c}{$\sigma$ ($N=60$)} \\
			\cline{2-11}
			$\delta$ & 95\% CL & 68\% CL & 95\% CL & 68\% CL & 95\% CL & 68\% CL & 95\% CL & 68\% CL & 95\% CL & 68\% CL \\
			\hline
			$10^{-2}$ & $[0, 0.31]$ & $[0,0.27]$ & $[0, 0.33]$ & $[0.19,  0.30]$ & $[0, 0.32]$ & $[0.09, 0.25]$ & $[0, 0.34]$ & $[0.20, 0.31]$ & $[0, 0.31]$ & - \\
			$10^{-3}$ & $[0, 0.61]$ & $[0, 0.53]$ & $[0, 0.64]$ & $[0.39, 0.59]$ & $[0, 0.61]$ & $[0.20, 0.48]$ & $[0, 0.66]$ & $[0.41, 0.60]$ & $[0, 0.62]$ & - \\
			$10^{-4}$ & $[0, 1.17]$ & $[0, 1.01]$ & $[0, 1.24]$ & $[0.76, 1.14]$ & $[0, 1.19]$ & $[0.42, 0.92]$ & $[0, 1.28]$ & $[0.80, 1.15]$ & $[0, 1.19]$ & - \\
			$10^{-5}$ & $[0, 3]$ & $[0, 2.09]$ & $[0, 3]$ & $[1.50, 2.94]$ & $[0, 3]$ & $[0.84, 1.85]$ & $[0, 3]$ & $[1.58, 2.93]$ & $[0, 3]$ & - \\
			\hline\hline
		\end{tabular}
	}
\end{table}

\begin{figure}[H]
	\centering
	\begin{subfigure}{0.47\textwidth}
		\centering
		\includegraphics[width=\textwidth]{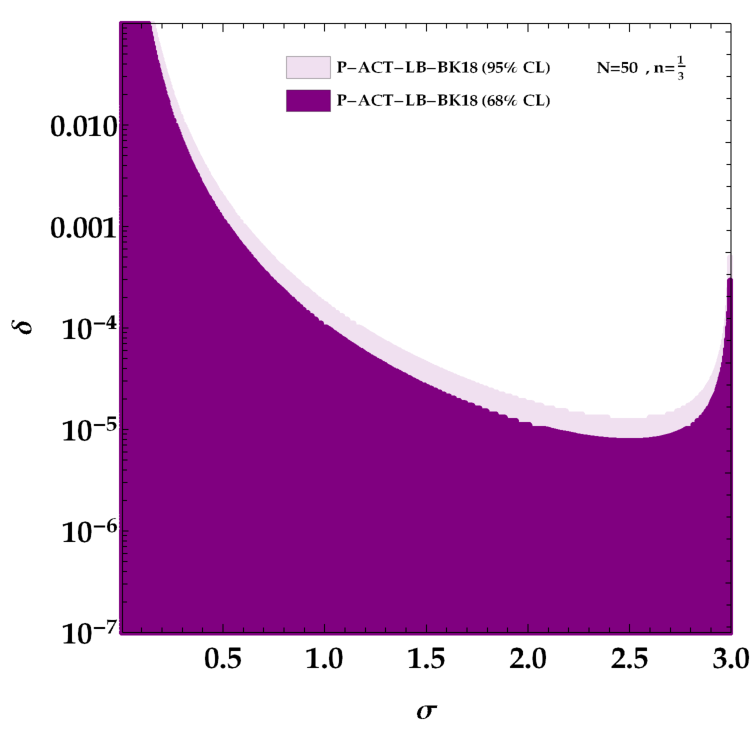}
		\label{fig-power-law_LQC5a}
		\caption{}
	\end{subfigure}
	\hfill
	\begin{subfigure}{0.47\textwidth}
		\centering
		\includegraphics[width=\textwidth]{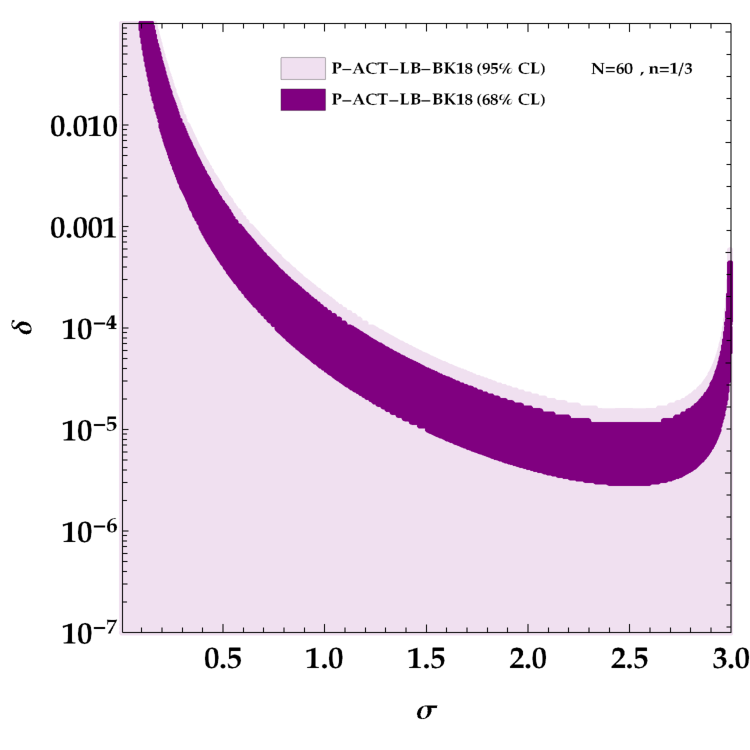}
		\label{fig-power-law_LQC5b}
		\caption{}
	\end{subfigure}
	\caption{Allowed phase space regions for the LQC inverse volume parameters ($\delta, \sigma$) corresponding to the potential $V(\varphi) \propto \varphi^{1/3}$ evaluated at (a) $N=50$ and (b) $N=60$ $e$-folds. The shaded regions denote the parameter space constrained by the joint P-ACT-LB-BK18 dataset at the 68\% (dark purple) and 95\% (light purple) CLs. }
	\label{fig-power-law_LQC5}
\end{figure}

\begin{figure}[H]
	\centering
	\begin{subfigure}{0.47\textwidth}
		\centering
		\includegraphics[width=\textwidth]{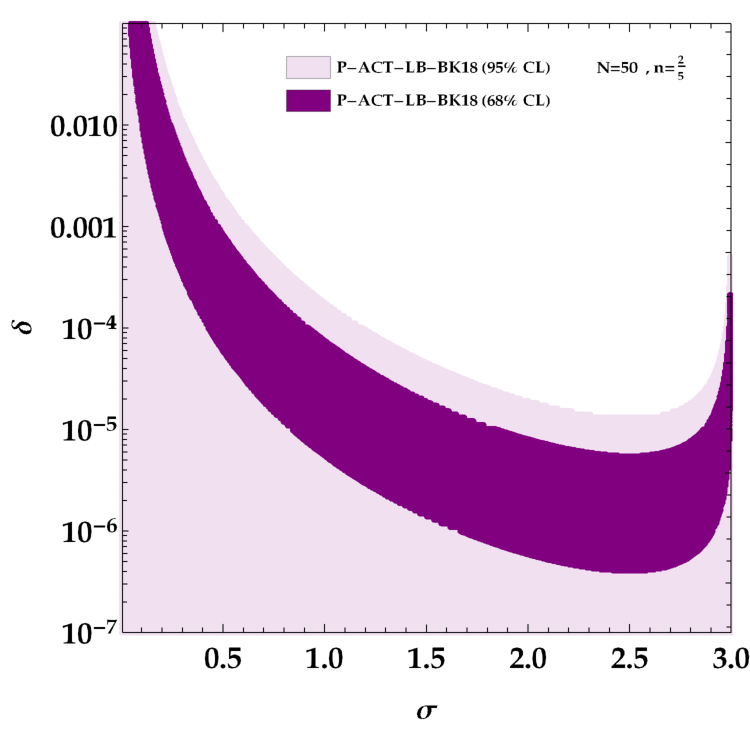}
		\label{fig-power-law_LQC4a}
		\caption{}
	\end{subfigure}
	\hfill
	\begin{subfigure}{0.47\textwidth}
		\centering
		\includegraphics[width=\textwidth]{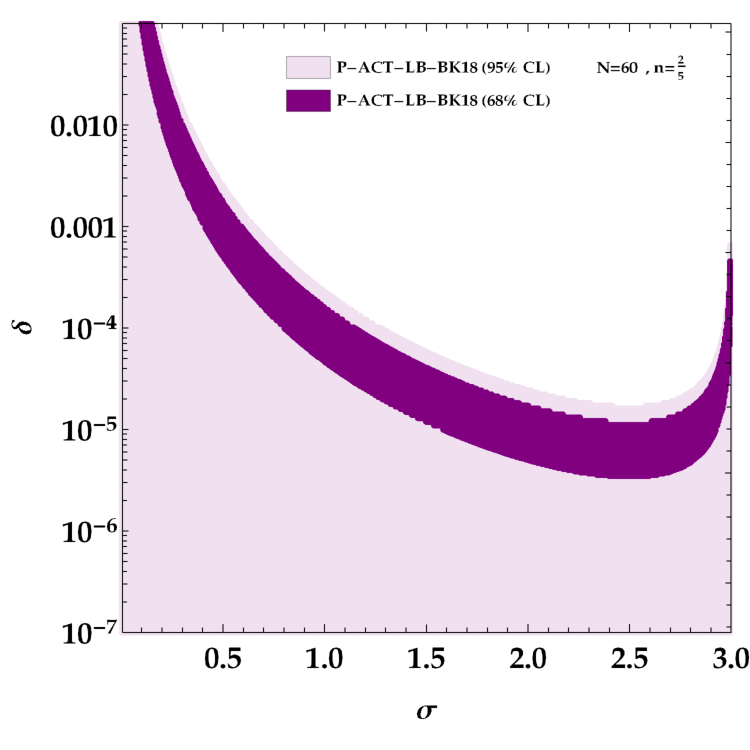}
		\label{fig-power-law_LQC4b}
		\caption{}
	\end{subfigure}
	\caption{Same as Fig. \ref{fig-power-law_LQC5}, but for the potential $V(\varphi) \propto \varphi^{2/5}$, evaluated at (a) $N=50$ and (b) $N=60$ $e$-folds.}
	\label{fig-power-law_LQC4}
\end{figure}

\begin{figure}[H]
	\centering
	\begin{subfigure}{0.47\textwidth}
		\centering
		\includegraphics[width=\textwidth]{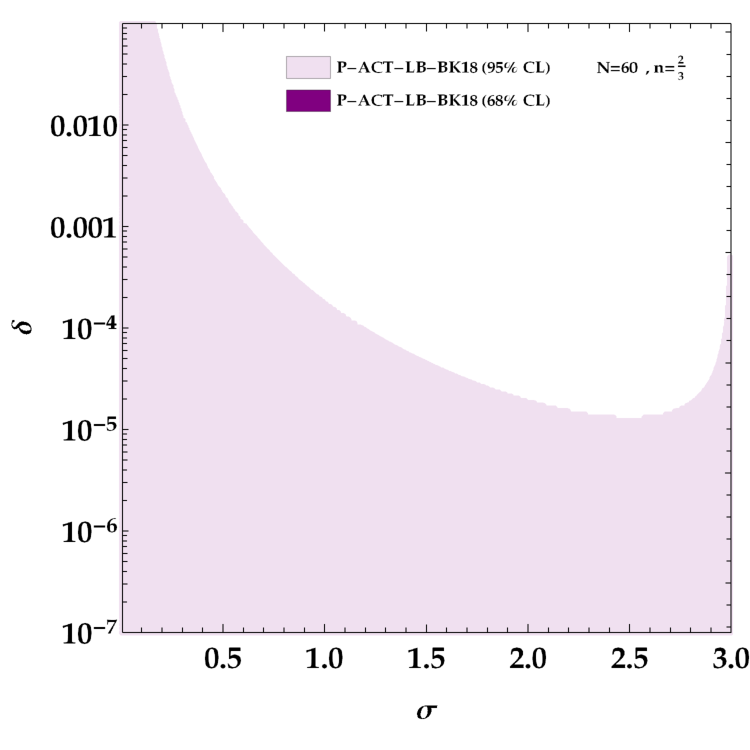}
		\captionsetup{labelformat=empty} 
		\caption{}
		\label{fig-power-law_LQC3a}
	\end{subfigure}
	\caption{Same as Fig. \ref{fig-power-law_LQC5}, but for the potential $V(\varphi) \propto \varphi^{2/3}$, evaluated at $N=60$ $e$-fold. Here, for the case of $N=50$, there is no region of parameter phase space consistent with the observations.}
	\label{fig-power-law_LQC3}
\end{figure}

Furthermore, we map the continuous allowed regions in the $\delta-\sigma$ parameter space through the phase space diagrams presented in Figs. \ref{fig-power-law_LQC5}, \ref{fig-power-law_LQC4}, and \ref{fig-power-law_LQC3} for $n=$1/3, 2/5 and 2/3, respectively. These plots reveal a distinct compensatory behavior between the two LQC parameters. Specifically, as the exponent $\sigma$ decreases, larger values of the amplitude $\delta$ are required to shift the predictions into the observationally favored regions, and vice versa. This compensatory mechanism highlights the robust nature of the inverse volume corrections in rescuing classically disfavored inflationary models, provided the parameters remain within the valid semi-classical regime.

\section{Conclusions}\label{sec:conclusion}
Here, we investigated the observational viability of the fractional power law inflationary potentials $V(\varphi) \propto \varphi^n$ $(n=1/3, 2/5, 2/3)$, within the effective framework of loop quantum cosmology incorporating inverse volume corrections. By evaluating the modified perturbation equations in the semi-classical regime and utilizing the uniform asymptotic approximation method, the corrected analytical expressions for the scalar spectral index $n_{\rm s}$ and the tensor-to-scalar ratio $r$ are obtained. Then, we estimated the predictions of the models in the $r-n_s$ plane and compared the obtained results with the latest high precision cosmological constraints, deduced from the joint analysis of ACT DR6, Planck 2018, DESI BAO, and BICEP/Keck 2018 datasets (P-ACT-LB-BK18).

In comparison with the standard classical inflation, the primary effect of LQC inverse volume corrections is a prominent negative shift in the scalar spectral index $n_{\rm s}$, while the impact on the tensor-to-scalar ratio $r$ remains comparatively sub-dominant. Consequently, increasing the LQC parameters including the amplitude $\delta$ and the exponent $\sigma$ translates the model predictions almost horizontally from right to left across the $r-n_{\rm s}$ plane. This robust translation mechanism successfully steers classically disfavored models back into the 95\% and 68\% CLs, thereby restoring their observational viability.

Finally, we systematically mapped the allowed regions of the $\delta-\sigma$ parameter space for each potential. Our phase space analysis revealed a compensatory relationship between the two quantum parameters: smaller values of the exponent $\sigma$ necessitate larger values of the amplitude $\delta$ to achieve the requisite shift into the observational bounds. Furthermore, the permitted parameter space is highly sensitive to the steepness of the potential. While the potential $V(\varphi) \propto \varphi^{1/3}$ accommodates a broad range of quantum geometric corrections, the allowed phase space shrinks significantly for steeper potentials, placing tight theoretical constraints on the inverse volume parameters for the potential $V(\varphi) \propto \varphi^{2/3}$. Ultimately, this study highlights that even strictly perturbative quantum gravity effects can leave distinct, observable imprints on primordial spectra, offering a compelling mechanism to reconcile simple inflationary models with precision cosmological data.

\subsection*{Acknowledgements}
The authors thank the referee for his/her valuable comments.

\appendix 
\section{}
\label{appendix:math_details}
This section provides the definitions of the quantities $Q_{-1}^{\star(s)}$, $Q_{0}^{\star(s)}$ and $Q_{1}^{\star(s)}$ appearing in Eq. (\ref{eq:power_spectrum_s1}), which are given by 
\begin{align}
	Q_{-1}^{\star(s)} &= \bar{Q}_{-1}^{(s)}, \nonumber \\
	Q_{0}^{\star(s)} &= (\sigma+2) \bar{Q}_{-1}^{(s)}\ln \left(\frac{3}{2}\right)+\bar{Q}_{0}^{(s)}, \nonumber \\
	Q_{1}^{\star(s)} &= 2\bar{Q}_{-1}^{(s)}\ln\left( \frac{3}{2}\right)+\bar{Q}_{1}^{(s)},
\end{align}
and
\begin{align}
	\bar{Q}_{-1}^{(s)} &= \frac{-\sigma^2(\sigma^2 - 2\sigma - 3)}{3}\alpha_0 L_1
	+ \frac{40\sigma^2(\sigma - 3)}{543}\alpha_0 L_3, \nonumber \\[6pt]
	\bar{Q}_{0}^{(s)} &= \left(\sigma^3 - \frac{\sigma^4}{3}\right)\alpha_0 L_2
	+ \left(\frac{40\sigma^3}{181} - \frac{40\sigma^4}{543}\right)\alpha_0 L_5
	+ \left(\frac{80\sigma^2}{181} - \frac{80\sigma^3}{543}\right)\alpha_0 L_3 \ln 2 \nonumber \\[6pt]
	&\quad + \left(\frac{40\sigma^4}{1629} + \frac{208\sigma^3}{1629} - \frac{368\sigma^2}{543}\right)\alpha_0 L_3
	+ \left(-\frac{2\sigma^4}{3} + \frac{4\sigma^3}{3} + 2\sigma^2\right)\alpha_0 L_1 \ln 2 \nonumber \\[6pt]
	&\quad + \left(\frac{\sigma^5}{9} + \frac{134\sigma^4}{543} - \frac{2614\sigma^3}{1629} - \frac{315\sigma^2}{181}\right)\alpha_0 L_1
	+ \left(-\frac{40\sigma^3}{1629} - \frac{40\sigma^2}{543} + \frac{80\sigma}{181}\right)\nu_0 L_3 \nonumber \\[6pt]
	&\quad + \left(-\frac{\sigma^4}{9} - \frac{4\sigma^3}{9} + \frac{5\sigma^2}{3} + 2\sigma\right)\nu_0 L_1
	- 3\sigma \chi L_1 + \frac{120\chi}{181} L_4, \nonumber \\[6pt]
	\bar{Q}_{1}^{(s)} &= \left(\sigma^2 - \frac{\sigma^3}{3}\right)\alpha_0 L_2
	+ \left(\frac{40\sigma^2}{181} - \frac{40\sigma^3}{543}\right)\alpha_0 L_5
	+ \left(\frac{40\sigma^2}{181} - \frac{40\sigma^3}{543}\right)\alpha_0 L_3 \ln 2 \nonumber \\[6pt]
	&\quad + \left(-\frac{40\sigma^4}{1629} + \frac{284\sigma^3}{1629} - \frac{68\sigma^2}{181} + \frac{40\sigma}{181}\right)\alpha_0 L_3
	+ \left(-\frac{\sigma^4}{3} + \frac{2\sigma^3}{3} + \sigma^2\right)\alpha_0 L_1 \ln 2 \nonumber \\[6pt]
	&\quad + \left(-\frac{\sigma^5}{9} + \frac{248\sigma^4}{543} - \frac{764\sigma^3}{1629} - \frac{20\sigma^2}{543}\right)\alpha_0 L_1,
\end{align}
with
\begin{equation}
	\chi \equiv \frac{\sigma
		\nu_0}{3} \left(\frac{\sigma}{6} + 1
	\right) + \frac{\alpha_0}{2}\left(5-\frac{\sigma}{3}
	\right).
\end{equation}
Furthermore
\begin{align}
	L_1 &\equiv \frac{\pi^{\frac{1}{2}}\Gamma(\frac{\sigma}{2})}{8\Gamma(\frac{\sigma+3}{2})}, \nonumber \\
	L_2 &\equiv \frac{\pi^{\frac{1}{2}}\Gamma(\frac{\sigma}{2})}{8\Gamma(\frac{\sigma+3}{2})}\Big[\psi^{(0)}\Big(\frac{\sigma}{2}\Big)-\psi^{(0)}\Big(\frac{\sigma+1}{2}\Big)\Big ], \nonumber \\
	L_3 &\equiv \frac{\pi^{\frac{1}{2}}\Gamma(1+\frac{\sigma}{2})(\sigma^2-3\sigma+2)}{24\Gamma(\frac{\sigma+1}{2})}, \nonumber \\
	L_4 &\equiv \frac{\pi^{\frac{1}{2}}(1+\sigma)\Gamma(1+\frac{\sigma}{2})(\sigma^2+2\sigma)}{48\Gamma(\frac{\sigma+3}{2})}, \nonumber \\
	L_5 &\equiv \frac{\pi^{\frac{1}{2}}\Gamma(\frac{\sigma}{2})}{48\sigma(\sigma-3)\Gamma(\frac{\sigma+3}{2})}\times\Big\{(\sigma^3-5\sigma^2+6\sigma)\times\nonumber \\
	&\Big[\psi^{(0)}\Big(\frac{\sigma-3}{2}\Big)-\psi^{(0)}\Big(\frac{\sigma}{2}\Big)\Big]-(2\sigma^2-12\sigma+12)\Big\}.
\end{align}
Here $\Gamma(x)$ and $\psi^{(0)}(x)$ are the Gamma and PolyGamma functions.



\end{document}